# Development of a time-to-digital converter ASIC for the upgrade of the ATLAS Monitored Drift Tube detector

*Jinhong Wang[b*], Yu Liang[a,b*], Xiong Xiao[b], Qi An[a], John W.Chapman[b], Tiesheng Dai[b], Bing Zhou[b], Junjie Zhu[b], Lei Zhao[a]*

[a] *State Key Laboratory of Particle Detection and Electronics, University of Science and Technology of China, Hefei, Anhui, 230026, PRC*
[b] *Department of Physics, University of Michigan, Ann Arbor, MI, 48109, USA*

ABSTRACT

The upgrade of the ATLAS muon spectrometer for high-luminosity LHC requires new trigger and readout electronics for the various elements of the detector. We present the design of a time-to-digital converter (TDC) ASIC prototype for the ATLAS Monitored Drift Tube (MDT) detector. The chip was fabricated in a GlobalFoundries 130 nm CMOS technology. Studies indicate that its timing and power consumption characteristics meet the design specifications, with a timing bin variation of ±40 ps for all 48 channels with a power consumption of about 6.5 mW per channel.

## 1. Introduction

### 1.1. Current MDT and Electronics

The ATLAS muon spectrometer [1] is composed of precision tracking and trigger chambers. Precision measurements of the muon track coordinates are provided by three stations of Monitored Drift Tube (MDT) chambers up to $|\eta|$ =2.5[1]. Cathode strip chambers (CSC) with higher granularity are used in the innermost station covering 2.0<$|\eta|$<2.7. Resistive plate chambers (RPC) located in the barrel region ($|\eta|$<1.05) and thin gap chambers (TGC) in the endcap region (1.05<$|\eta|$<2.4) are used to provide trigger information.

The MDT system is the main component of the ATLAS precision muon tracking system and is designed to provide a stand-alone muon transverse momentum measurement with a resolution of 10% for 1 TeV muons. There are, in total, 1,150 MDT chambers made from 354,000 aluminum tubes covering a total area of 5,500 $m^2$. The chambers have been operated successfully since 2009 and have made important contributions to the discovery of the Higgs boson, precision Standard Model measurements and exotic searches at ATLAS.

Each tube has an inside diameter of 29.97 mm and is filled with a mixture of $Ar/CO_2$ (93/7) at 3 bar. A 50 µm gold-plated tungsten wire is positioned at the center of each tube. A high voltage of 3.08 kV is imposed across the tube wall and the central wire. Ionization created by the passage of a muon track can take up to 750 ns to reach the anode wire.

The present MDT readout electronics [2] is designed to preserve the inherent measurement accuracy of the tubes (80 um) and to cope with the high hit rates expected at the full LHC luminosity. Raw signals from 24 tubes are routed via signal hedgehog boards and processed by three custom-designed monolithic Amplifier/Shaper/ Discriminator (ASD) chips [3]. The ASD differential output signals are then routed to a Time-to-Digital Converter (TDC) chip [4] where the arrival times of leading and trailing edges are digitized and stored in a buffer memory waiting for the ATLAS first-level trigger accept signal. Each frontend mezzanine card contains three 8-channel ASDs and one 24-channel TDC. Up to 18 mezzanine cards are controlled and readout by an FPGA on the Chamber Service Module (CSM). The CSM communicates with the off-chamber electronics via two fibers, one coming from the Timing, Trigger and Control distribution box [5] and the other going to the MDT Readout Driver (MROD) [6].

---

*\* These authors contributed equally to this work.*

[1] ATLAS uses a right-handed coordinate system with its origin at the nominal interaction point in the center of the detector and the z-axis along the beam pipe. The x-axis points from the IP to the center of the LHC ring, and the y-axis points upward. The pseudorapidity is defined in terms of the polar angle $\theta$ as $\eta = - \ln \tan(\theta/2)$.

*1.2. MDT frontend electronics for high-luminosity LHC (HL-LHC)*

The LHC collider has a few upgrades scheduled in the years 2019-2024 with the instantaneous luminosity increased to $5-7 \times 10^{34}$ cm$^{-2}$s$^{-1}$. The machine is expected to collect about 3000 fb$^{-1}$ of data by 2035. ATLAS detector will have corresponding upgrades to its trigger and readout systems to handle the large data rates and demanding environment imposed by these LHC upgrades.

To handle large number of background hits and low momentum muon tracks expected at the trigger level, ATLAS plans to include MDT chamber data in the first trigger level to provide additional rejection of fake muons and to reject low momentum muons by sharpening the trigger turn-on curve. The design of the MDT on-chamber electronics system for the HL-LHC is based on sending all muon hits off chamber to new trigger and readout circuitry in the ATLAS counter room. In this trigger-less scheme, all raw tube signals are amplified, shaped, discriminated, and digitized on-chamber and sent via the CSM to a Hit Extraction Board (HEB). Due to the long drift time of the MDT signal, the timing information from the fast trigger chambers (RPCs in the barrel and TGCs in the endcap) are used as the muon reference time. The fast trigger chamber reference time is used to select possible MDT candidate hits and to convert the observed times to a reduced resolution radial positon in the tubes. The radial position information for all matched hits are sent to the MDT trigger processor where the segment-finding and track-fitting algorithms are performed, while the full-resolution time information for all matched hits are stored for transmission to FELIX [7] after receiving the first-level trigger accept signal. Simulations have demonstrated that this design meets desired performance with achievable increases in data bandwidth.

Figure 1 shows the overall schematic diagram for the MDT trigger and readout system at HL-LHC. The TDC is a crucial component in the new trigger architecture. It is responsible for the time digitization of raw tube signals, which is the basis for all following trigger processing. Although the existing TDC [4] has a trigger-less mode, it cannot meet the requirement at the HL-LHC due to the increased data rate. Moreover, the new TDC chip must also have significantly reduced legacy [8]. This paper focuses on the design of a new TDC targeted for the ATLAS MDT chambers at the HL-LHC.

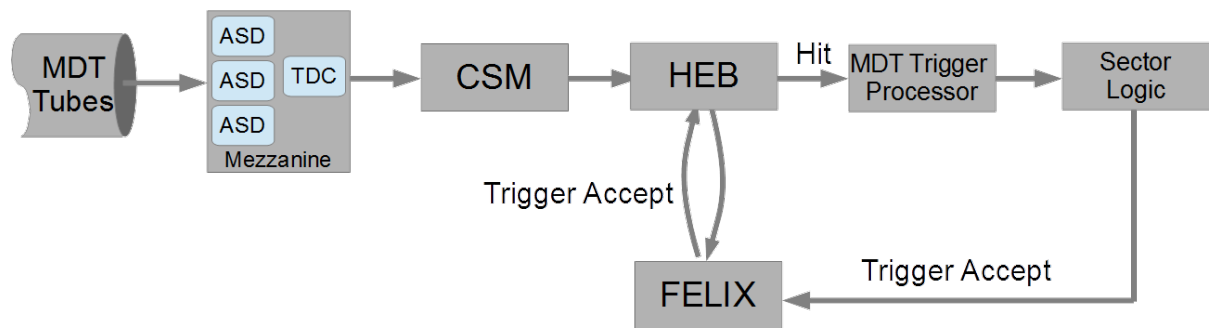

**Fig. 1: Schematic diagram of the MDT trigger and readout system at the HL-LHC.**

## 2. Design of a trigger-less TDC for the ATLAS MDT upgrade

In the MDT trigger and readout system, raw signals from groups of 24 tubes are processed by three custom-designed ASD ASICs, from which the timing information is extracted in the form of "Time-Over-Threshold" (TOT) pulses. A single TDC is responsible for time digitization of both leading and trailing edges of the TOT pulses. A ~780 ps time resolution is required over a dynamic range of one LHC orbit cycle (~102.4 us). This corresponds to a 17-bit time measurement. In addition, the power consumption of one TDC chip needs to be less than 350 mW as no active cooling system is available.

To meet the timing requirement, we break the time measurement into coarse-time and fine-time measurements, where the dynamic range is covered by the coarse-time measurement and the resolution is achieved from the fine-time measurement. The range of the fine-time measurement is one coarse-time step. The 17-bit time is then formed from the coarse-time bits plus the fine-time bits. It is possible to derive the fine time from multiple phases to achieve the required time resolution while avoiding high frequency clocking. We implement the circuit by using the TDC input signal to sample several phase-shifted clocks, as shown in Fig. 2(a). In contrast to sampling hits with multiple phases (as illustrated in Fig. 2(b)), the proposed architecture is simpler, requires less logic, and consumes less power. This is because the samples from the implementation in Fig. 2(b) are in different clock domains and additional circuit is needed to migrate the samples into the same domain for signal processing.

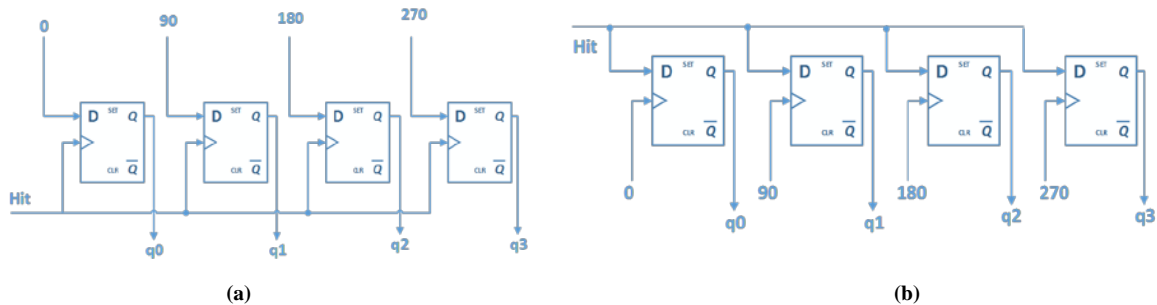

**Fig. 2 Implementation of the fine time measurement of the TDC – (a) Hit samples clocks; (b) clocks sample Hit.**

The clock phase interval chosen is ~780 ps and the number of phases is determined by the operating clock frequency used for the coarse-time measurement. A higher frequency results in fewer phases for the fine-time interpolation whereas the number of bits for the coarse time will be greater to cover the dynamic range. The choice made for the 130 nm CMOS technology is 320 MHz so that only four phases are necessary, i.e. 0°/90°/180°/270°. Since the phases differ by 90 degrees, in principle two clocks with a phase shift of 90° is adequate if we make use of both rising and falling edges. The bit width for fine-time and coarse-time measurements are thus 2 bits and 15 bits respectively. A 50% duty cycle of the clock is required for timing precision. The final time measurement of a hit is a combination of the coarse-time and fine-time measurements.

## 2.1. Implementation of the Fine-Time Measurement

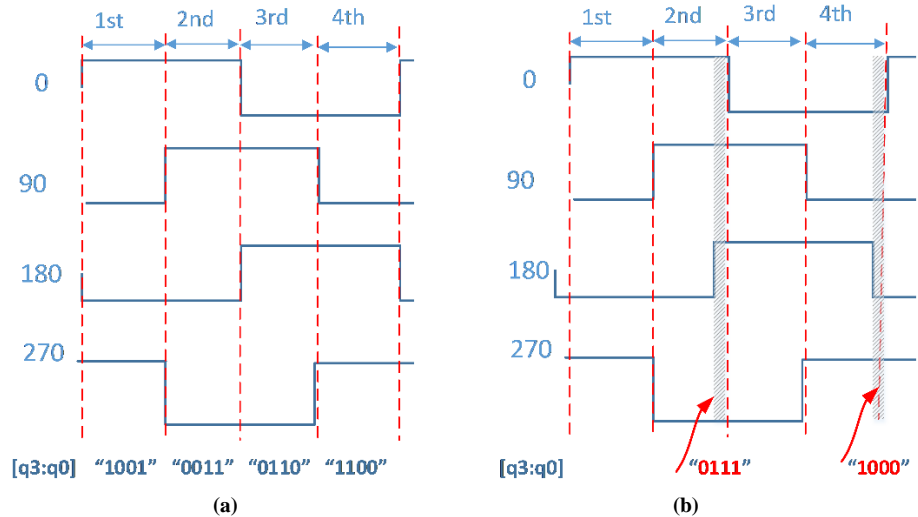

**Fig. 3: Encode of the TDC fine-time measurement – (a) Ideal case; (b) practical case with un-aligned clock phases.**

For the fine-time implementation as shown in Fig. 2(a), the four 320 MHz clocks are fed into the data inputs of four sampling registers. Once a hit signal presents, the status of the current phases of the four clocks are sampled as q3-q0, from which the fine-time measurement of the hit signal is decoded as 2 bits. For the ideal case, there are only four possible combinations of q[3:0], as shown in Fig. 3(a). Whereas in reality, other combinations are also possible due to the sampling meta-stability or ambiguous phase relationships. Meta-stability of sampling registers occurs upon setup/hold time failures. The hit signals are asynchronous to the 320 MHz clocks, thus sampling meta-stability could certainly happen on occasions. Ambiguous phase relationships present when 0°/180° or 90°/270° pair clocks are not exactly reversed in phase due to non-uniform delay induced in the clock routing or buffering. One example is given in Fig. 3(b), in which the asymmetry of the 0°/180° pair introduces gaps as dashed bands. A hit signal that arrives within the gaps gives a q[3:0] that is different from the ideal one, e.g., the gap between the 2nd and 3rd phase gives a code "0111", which is not any of the four ideal codes.

Both sampling meta-stability and ambiguous phases introduce exceptional combinations of q[3:0], and their contribution is indistinguishable in view of the status of q[3:0]. We analyze and summarize all possible q[3:0] combinations as in Table 1 for each edge, e.g. edge 0 corresponds to the edge at the time of the rising of the 00 clock. Edges 1-3 are counted from edge 0 sequentially. For edge 0, q1 and q3 are stable while q0 and q2 are vulnerable. As a result, there are four possible combinations at edge 0, with "1001" as the idea case and "1101", "1000" and "1100" are the exceptions. Correction can be made knowing that q1 and q3 are fixed at value 0 and 1 respectively.

A summary of the corrections are list on the right side of Tab. 1. For the three exceptions at edge 0, "1101" and "1000" are recovered correctly, while "1100" collides with the ideal case of edge 3. The fine-time code increases by 3 steps in this scenario. However, in view of the coarse-time selection in the following section, the coarse time is backward by 1 count, which is equivalent to reduce 4 fine steps. As a result, the final time measurement only varies by 1 fine step from the ideal case. Similar analysis can be done for the other three edges and the results are summarized in Tab. 1. There are only 12 combinations of q[3:0] as listed and the remaining four "1111", "0000", "0101" and "1010" will never happen due to the reverse relation between q0/q2 and q1/q3. All possible exceptions can be corrected and the error induced is no more than 1 fine bin size (~780 ps).

**edge0**

|      | q0 | q1 | q2 | q3 |
|------|----|----|----|----|
| ideal | 1 | 0 | 0 | 1 |
| meta | 1 | 0 | 1 | 1 |
| meta | 0 | 0 | 0 | 1 |
| meta | 0 | 0 | 1 | 1 |

q[3:0]

| raw | correction | encoded | coarse |
|-----|------------|---------|--------|
| 1001 | 1001 | 0 | 0 |
| 1101 | 1001 | 0 | 0 |
| 1000 | 1001 | 0 | 0 |
| 1100 | 1100 | 3 | -1 |

**edge1**

|      | q0 | q1 | q2 | q3 |
|------|----|----|----|----|
| ideal | 1 | 1 | 0 | 0 |
| meta | 1 | 1 | 0 | 1 |
| meta | 1 | 0 | 0 | 0 |
| meta | 1 | 0 | 0 | 1 |

q[3:0]

| raw | correction | encoded | coarse |
|-----|------------|---------|--------|
| 0011 | 0011 | 1 | 0 |
| 1011 | 0011 | 1 | 0 |
| 0001 | 0011 | 1 | 0 |
| 1001 | 1001 | 0 | 0 |

edge2

|      | q0 | q1 | q2 | q3 |
|------|----|----|----|----|
| ideal | 0 | 1 | 1 | 0 |
| meta | 0 | 1 | 0 | 0 |
| meta | 1 | 1 | 1 | 0 |
| meta | 1 | 1 | 0 | 0 |

q[3:0]

| raw | correction | encoded | coarse |
|-----|------------|---------|--------|
| 0110 | 0110 | 2 | 0 |
| 0010 | 0110 | 2 | 0 |
| 0111 | 0110 | 2 | 0 |
| 0011 | 0011 | 1 | 0 |

**edge3**

|      | q0 | q1 | q2 | q3 |
|------|----|----|----|----|
| ideal | 0 | 0 | 1 | 1 |
| meta | 0 | 0 | 1 | 0 |
| meta | 0 | 1 | 1 | 1 |
| meta | 0 | 1 | 1 | 0 |

q[3:0]

| raw | correction | encoded | coarse |
|-----|------------|---------|--------|
| 1100 | 1100 | 3 | 0 |
| 0100 | 1100 | 3 | 0 |
| 1110 | 1100 | 3 | 0 |
| 0110 | 0110 | 2 | 0 |

**Table 1: TDC fine-time encoding and correction scheme.**

The occurrence probability of these exceptional combinations is proportional to the setup/hold time of the sampling registers, and also to the "gap" induced from ambiguous phases. Proper choice of sampling registers with shorter setup/hold time and careful arrangement of the clock reversion and fanout distribution can both lower the possibility of these exceptional bits. In our implementation, we choose a sampling register with a typical sum of setup and hold time of ~20 ps. Identical clock distributions are applied for all four clocks and compensations are performed to minimize the distortions such as uneven high/low ratio of clocks and mis-alignment of clock edges.

*2.2. Implementation of the Coarse-Time Measurement*

The coarse time of the TDC is derived from coarse counters, as shown in Fig. 4. Two 15-bit binary counters (CNT1 and CNT2) running at the rising and failing edges of a replica of the 320 MHz clock with a phase of 0°, respectively [9]. Anytime a Hit signal arrives, both counters are sampled, and the final coarse time is selected away from transitions based on the value of the fine time measurement. In the design, CNT2 is configured to run earlier than CNT1 as shown in Fig. 4. For fine time 0-1, CNT2 is selected; otherwise, CNT1 is selected. The dual counter scheme is to avoid the uncertainty region of counter value switching, shown as the dashed regions in the Fig. 4.

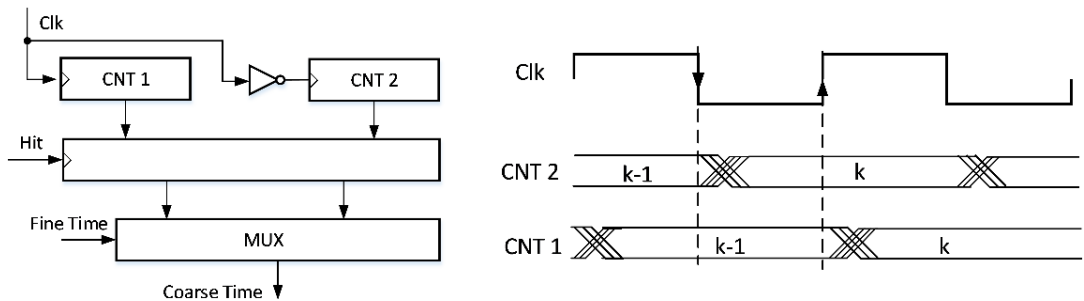

**Fig. 4: The TDC coarse time counters.**

*2.3. Implementation of the TDC ASIC*

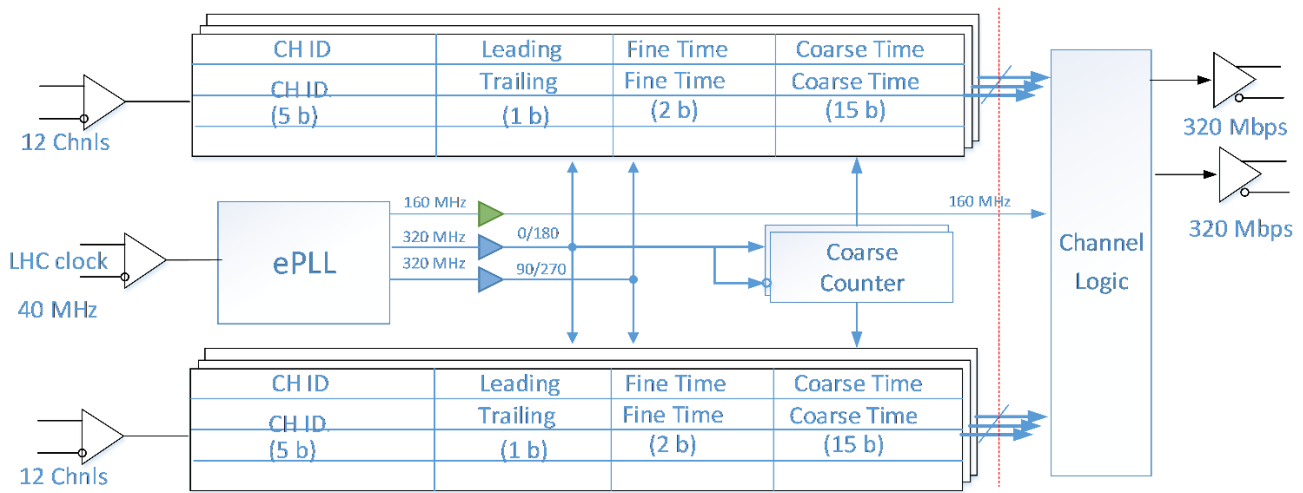

**Fig. 5: The block diagram of the TDC ASIC.**

A block diagram of the TDC is shown in Fig. 5, in which the two phase shifted clocks are obtained from the CERN ePLL [10] and the 24 TDC channels are placed symmetrically on the two sides, with 12 channels each. The reference clock to the ePLL is the LHC bunch crossing clock, which is about 40 MHz. The "channel logic" block accepts all raw time measurements from each of the TDC channels. It performs time calibration and edge pairing etc., depending on the TDC operation mode. The time measurements from each channel are buffered in separate FIFOs and are read out via a priority encoding of the FIFO status. There are two output ports, each handling 12 channels. The TDC output rate is either 160 Mbps or 320 Mbps and is determined by the maximum bandwidth used on the output cables.

The fine-time and coarse-time units comprises a complete rising-edge-sensitive TDC block. For the TOT operation of the TDC, another replica of the TDC block is utilized for the trailing edge of a TOT signal. Together, the two blocks make up a full

TDC channel for the MDT detector. The usage of a separate TDC block to process the trailing edge of TOT signals relaxes the restriction on the TOT pulse width.

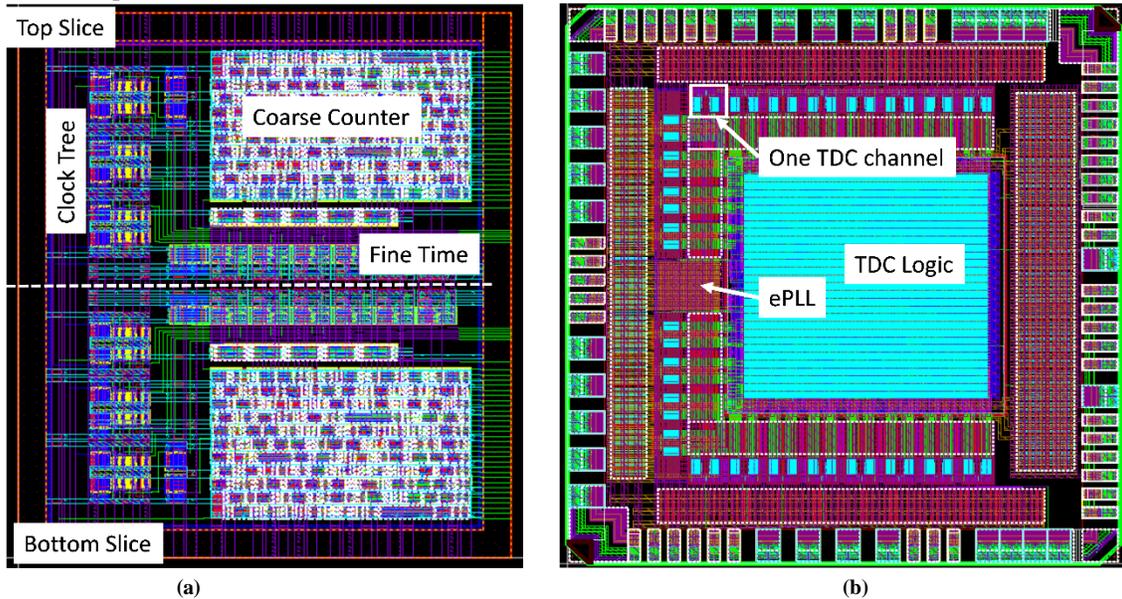

**Fig. 6: (a) Layout of a single TDC channel with dual edges (with an area of 185 um × 177 um); (b) layout of the whole TDC ASIC.**

A picture of a single-channel TDC layout is shown in Fig. 6(a), in which two identical TDC slices are utilized to process both edges of a TOT signal. Since all 24 channels are independent of each other, the whole implementation is a replica of the single-channel module which results in 48 TDC slices in total.

## 3. Performance of the new TDC ASIC

A first TDC prototype has been designed, fabricated and tested. Fig. 6(b) shows the floorplan of this ASIC. The area of a TDC slice is 195 um × 177 μm, and the total area of the chip is 3.7 × 3.4 mm$^2$. The chip was fabricated in a GlobalFoundries 130 nm CMOS process and is packaged using a QFN100 package.

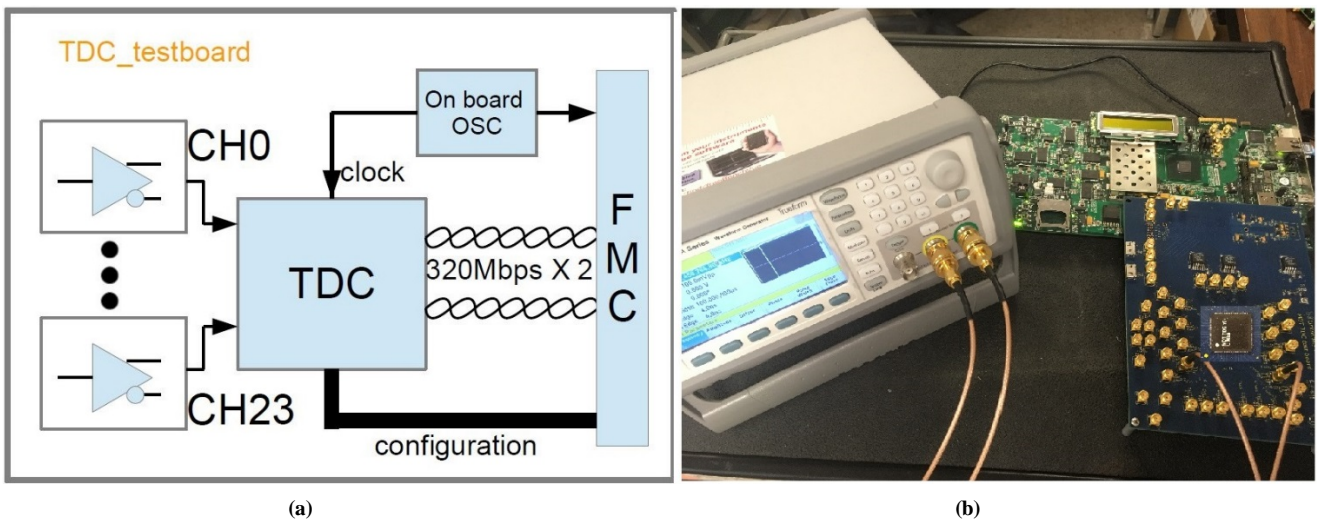

**Fig. 7: (a) A diagram of the TDC test board; (b) setup for the TDC performance test.**

A test board has been designed and fabricated to evaluate the TDC performance. A block diagram of the TDC test board is shown in Fig. 7(a), in which single-end hit signals are converted into LVDS signals by a high-speed low-noise discriminator (ADCMP604) before being fed into the TDC. The TDC performs time digitization and the measured time information are shifted out via two lines at 320 Mbps to an field-programmable gate array (FPGA) evaluation board through an FPGA mezzanine card (FMC) connector. A high-precision on-board oscillator (OSC) provides a reference clock to the TDC for the time digitization and to the FPGA for TDC data decoding. A picture of the TDC test system is shown in Fig. 7(b). Inputs to the TDC are provided by a signal generator with programmable time intervals. A Xilinx FPGA evaluation board collects and sends the TDC data to a computer via an Ethernet port for further data analysis.

The fine-time bin size is evaluated with the code-density test [11]. A typical distribution of the raw bins (q [3:0]) in a TDC slice is shown in Fig. 8(a) together with four ambiguous combinations ("1110", "0100", "1011" and "1101"). After on-line corrections as described in Table 1, only four time bins are present in the TDC fine-time measurement. Corresponding differential nonlinearity (DNL) of these four bins is shown in Fig. 8(b).

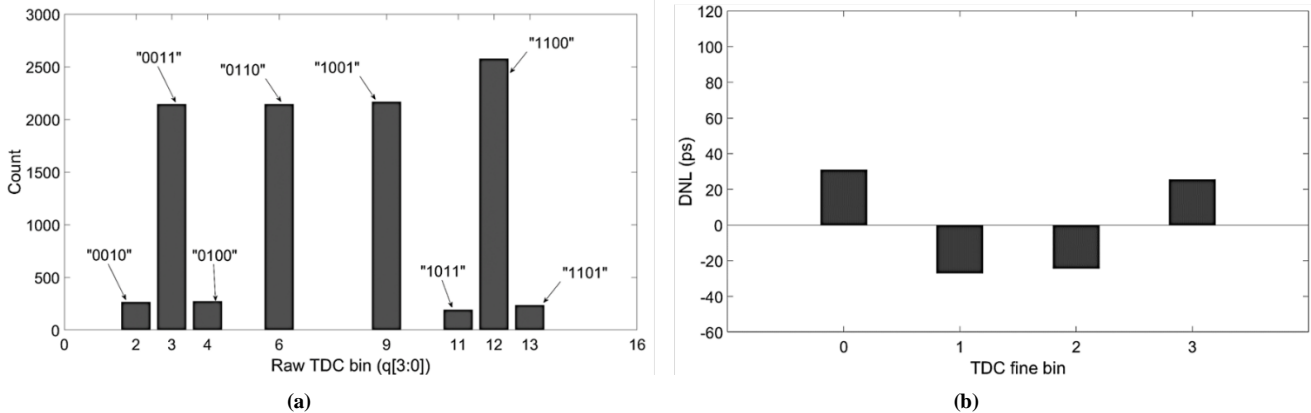

**Fig.8: (a) Distribution of raw bits (q[3:0]) in a TDC slice; (b) differential nonlinearity after corrections.**

Figure 9 shows the bin size for four bins in all 48 TDC slices. The horizontal axis is the TDC slice number (1-48), whereas the vertical axis is the size of the four bins in a slice. The mean size is found to be 781 ps with a variation ±40 ps, i.e. ±5% of the bin size. Figure 10 summarizes the DNL and integral nonlinearity (INL) of all 48 slices and they are found to be within ±40 ps.

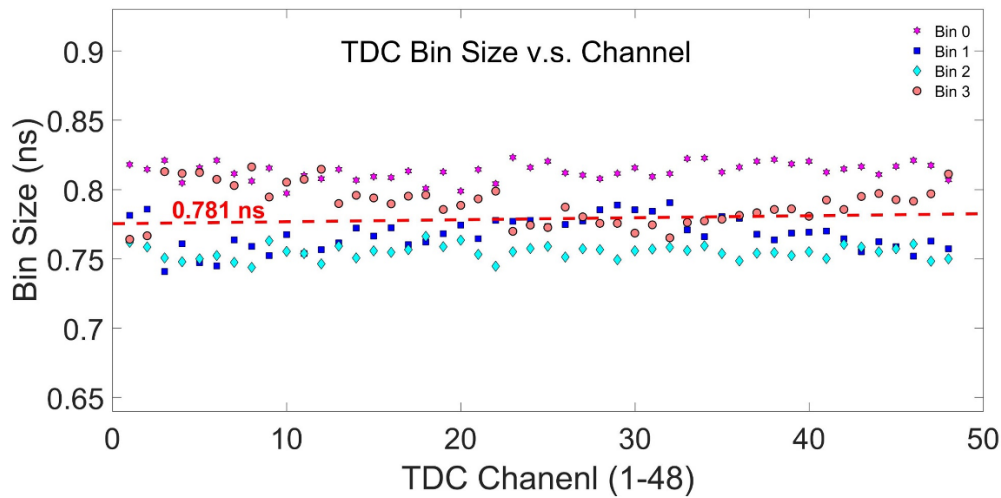

**Fig. 9: Fine-time bin size for all 48 slices.**

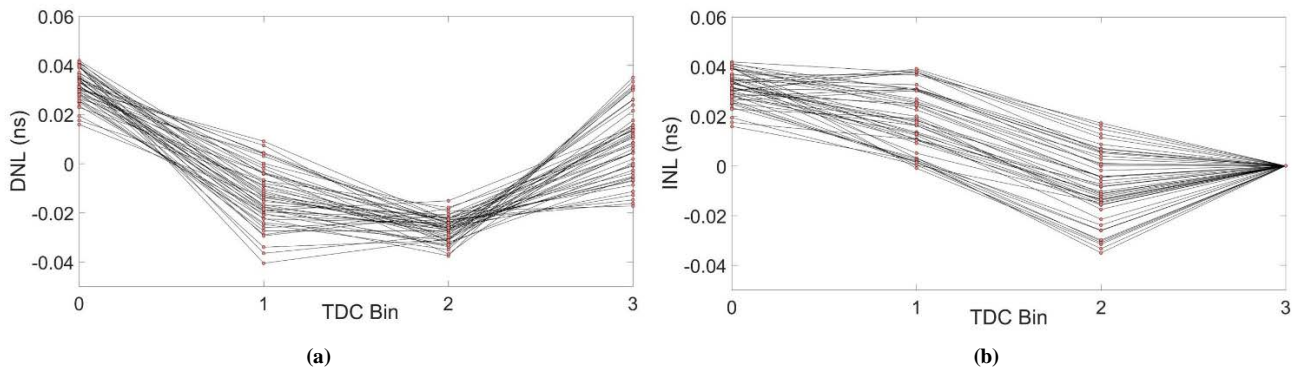

**Fig. 10: (a) DNL and (b) INL of all 48 TDC slices**

We swept the timing performance of each TDC slice for different time intervals. A typical result for one slice is shown in Fig. 11. The horizontal axis is the time interval measured by the TDC, and the vertical axis is the RMS timing precision in unit of a TDC bin (LSB: Least Significant Bit, 781 ps). The result agrees with the prediction of the time resolution induced solely by the quantization of the TDC fine-time bin size [12], which indicates that the contribution to the overall time resolution from other sources is negligible and the timing performance of the TDC is close to the theoretical prediction.

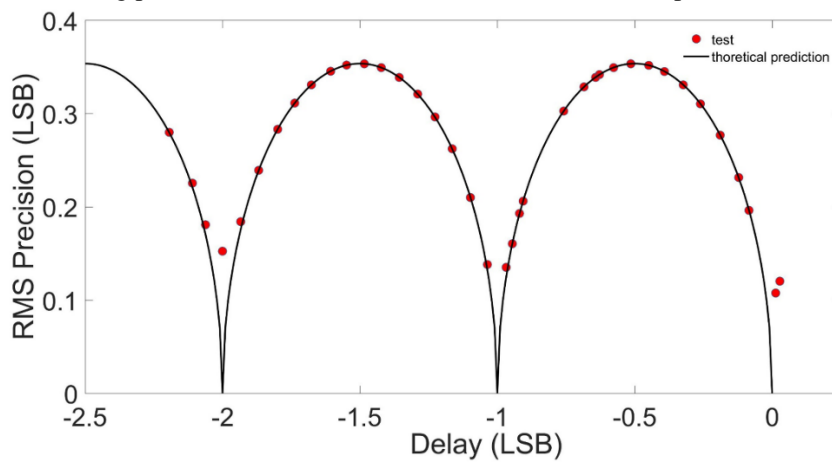

**Fig. 11: Timing precision of a TDC slice with respect to different time intervals.**

We also check the timing performance using cable delay measurements [13]. The input signals to two channels are delayed by a known time interval, and the difference of the digitized times for these two channels in a unit of 781 ps is checked. Fig. 13 shows the digitized time difference for two time intervals. Figure 12(a) presents the time difference with an input time interval of 0.5 ps, the distribution has a RMS of 36 ps; while Fig. 12(b) shows the time difference with an input time interval of 401 ps (close to half of an LSB), the distribution has a RMS of 275 ps. The timing uncertainty is dominated by the uncertainty due to the quantization of the TDC fine-time bin size.

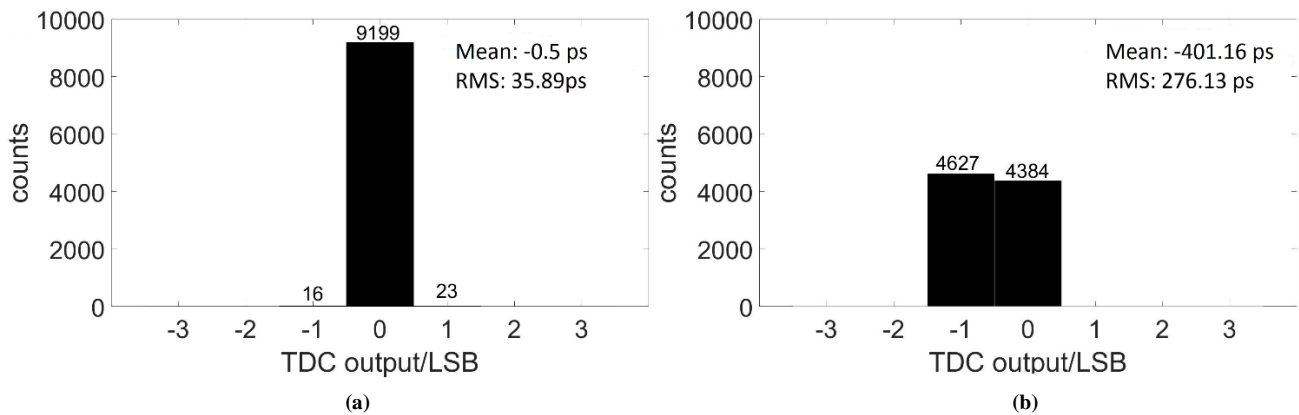

**Fig. 12: (a) Code distribution for a time interval of 0.5 ps; (b) code distribution for a time interval of 401.16 ps.**

The power consumption of the TDC is measured by connecting a 0.1 Ω resistor in serial with the power supply and monitoring the current across the resistor. Test shows that the total current is ~206 mA, corresponding to a total power consumption of 310 mW with a supply of 1.5 V. All 48 TDC slices are running when we perform the measurement, thus the average power consumption per TDC slice is ~6.5 mW.

**4. Summary**

A prototype TDC ASIC has been designed, fabricated, and tested for the upgrade of the ATLAS MDT detector at the HL-LHC. A detailed design was presented that meets the required timing precision and power consumption. Tests were performed and the performance of the TDC has been shown to meet the design specifications for the HL-LHC. The TDC shows a good channel uniformity of 5% of the bin size over 48 TDC slices. Further developments of this ASIC will be focused on adding an additional option to allow the ASIC to run in a triggered mode and only send out the timing information after receiving the first-level trigger accept signal. This addition will be implemented in the "TDC logic" part. We expect to finalize the design in 2020 and have chips certified and installed on chambers around 2024.

**Acknowledgments**

This work is supported by National Key Program for S &T Research and Development (Grant NO.: 2016YFA0400100) and the US Department of Energy under contracts DESC0008062 and DE-AC02-98CH10886.